\documentclass[onecolumn,preprintnumbers,amssymb,preprint,nofootinbib]{revtex4}
\usepackage[utf8]{inputenc}
\usepackage[english]{babel}
\usepackage[T1]{fontenc}
\usepackage{graphicx,amssymb,amsmath,wasysym}
\usepackage[bookmarks=false]{hyperref}
\usepackage{graphicx}
\usepackage{dcolumn}
\usepackage{bm}

\newcommand{\be}[1]{\begin{equation} \label{(#1)}}
\newcommand{\ee}{\end{equation}}
\newcommand{\ba}[1]{\begin{eqnarray} \label{(#1)}}
\newcommand{\ea}{\end{eqnarray}}
\newcommand{\rf}[1]{(\ref{(#1)})}

\newcommand{\Eres}{E_{\rm r}}
\newcommand{\mn}{m_N}
\newcommand{\mphi}{m_\phi}
\newcommand{\mdm}{m_{\rm DM}}
\newcommand{\mr}{m_{\rm r}}
\newcommand{\zr}{z_{\rm r}}
\newcommand{\gr}{\Gamma_{\rm r}}

\begin{document}

\input{epsf}

\def \a'{\alpha'}
\baselineskip 0.65 cm
\begin{flushright}
IFIC/14-02 \\
\end{flushright}

\begin{center}{\large
    {\bf }}
  {\vskip 0.5 cm}

  \begin{center}
    \textbf{{\Large  Dips in the Diffuse Supernova Neutrino
        Background} }\\

    \vspace{0.5in}

    \textbf{$\textrm{Yasaman
        Farzan}^a$~\footnote{yasaman@theory.ipm.ac.ir} and
      $\textrm{Sergio
   Palomares-Ruiz}^{b}$~\footnote{Sergio.Palomares.Ruiz@ific.uv.es}}
    \\
    \vspace{0.2in}
    \textsl{${}^a$ School of physics, Institute for Research in
      Fundamental Sciences (IPM),\\ P.O.Box 19395-5531, Tehran, Iran}\\
    \vspace{0.2in}
    \textsl{${}^b$ Instituto de F\'{\i}sica Corpuscular (IFIC),
      CSIC-Universitat de Val\`encia,\\ Apartado de Correos 22085,
      E-46071 Valencia, Spain} \\
    \vspace{.75in}
  \end{center}

\end{center}

\begin{abstract}
Scalar (fermion) dark matter with mass in the MeV range coupled to
ordinary neutrinos and another fermion (scalar) is motivated by
scenarios that establish a link between radiatively generated neutrino
masses and the dark matter relic density.  With such a coupling,
cosmic supernova neutrinos, on their way to us, could resonantly
interact with the background dark matter particles, giving rise to a
dip in their redshift-integrated spectra.  Current and future
neutrino detectors, such as Super-Kamiokande, LENA and
Hyper-Kamiokande, could be able to detect this distortion.
\end{abstract}

\maketitle
\newpage

\section{Introduction}
\label{sec:introduction}

Core-collapse supernova (SN) explosions of  type II, Ib and Ic are
known sources of neutrinos with energies in the range of few to tens
of MeV.  If a SN explosion occurs  in our galaxy at a distance of
10~kpc, over $10^4$ events can be observed in current detectors such
as Super-Kamiokande (SK)~\cite{Ikeda:2007sa} or the proposed detectors
such as LENA~\cite{Wurm:2011zn} as well as in the liquid scintillator
detector to be installed in the ANDES
observatory~\cite{Machado:2012ee}.   In addition, the IceCube neutrino
telescope would register about $10^6$ photons in excess of its
background.   Moreover,  a number of smaller existing and upcoming
detectors can collect tens to hundreds of SN
events~\cite{Scholberg:2007nu, Scholberg:2010zz}. However, galactic
core-collapse SN events are rare~\cite{Ikeda:2007sa, Alekseev:2002ji}
and  many years might pass before registering such an explosion in
our galaxy.

On the other hand, the diffuse SN neutrino background (DSNB), from
all the SN explosions that have occurred in the history of the
Universe, is a guaranteed flux.  The DSNB flux depends on the cosmic
star formation rate~\cite{Hopkins:2006bw, Horiuchi:2008jz,
  Horiuchi:2011zz}, which is relatively well known up to redshift $z
\simeq 9$~\cite{Hopkins:2006bw, Kistler:2013jza}.  Although so far
there is only an upper bound on this diffuse neutrino
flux~\cite{Malek:2002ns, Lunardini:2008xd, Bays:2011si, Baysthesis},
the limits are very close to the expectations from recent theoretical
predictions~\cite{Totani:1995rg, Totani:1995dw, Malaney:1996ar,
  Hartmann:1997qe, Totani:1997vj, Kaplinghat:1999xi, Ando:2002ky,
  Fukugita:2002qw, Strigari:2003ig, Ando:2004sb, Ando:2004hc,
  Iocco:2004wd, Strigari:2005hu, Lunardini:2005jf, Daigne:2005xi,
  Horiuchi:2008jz, Lunardini:2009ya, Lunardini:2012ne,
  Tamborra:2012ac}.  The proposed Gadollinium-doped SK
phase~\cite{Beacom:2003nk}, currently under study within the EGADS
project~\cite{Watanabe:2008ru}, and the LENA
detector~\cite{Wurm:2011zn} could detect up to $\sim$10 $\bar{\nu}_e$
events per year.  The proposed Hyper-Kamiokande (HK) detector, with a
fiducial volume 25 times larger than SK~\cite{Abe:2011ts}, will be
able to detect $\sim$200 events per year.  If the IceCube extension
MICA is ever built~\cite{Cowen:2013}, with a volume of $\sim$5~Mton
and a low energy threshold,  ${\cal O}(1000)$ events per year could be
collected.  Thus, it is in principle possible to foresee that we
will be able to reconstruct the $\bar{\nu}_e$ energy spectrum of the
DSNB, which not only would allow us to constrain the SN models, but
also to search for unexpected surprises.

As it is well known~\cite{Weiler:1982qy, Weiler:1983xx, Roulet:1992pz,
  Yoshida:1996ie, Eberle:2004ua, Barenboim:2004di}, the spectrum of
neutrinos with extremely high energies (i.e., $E_\nu = m_Z^2/(2
m_\nu)>10^{21}$~eV) could be distorted by the resonant interaction off
cosmic relic neutrinos at the $Z$-pole.  In a similar fashion,
neutrinos from cosmological core-collapse SN events may interact on
their way to Earth with the intergalactic matter leading to a
deformation of their spectrum.  For instance, in models with
additional light $Z'$ gauge bosons coupled to neutrinos, SN neutrinos
could interact with the low energy relic neutrino background giving
rise to a dip in their spectrum~\cite{Goldberg:2005yw,
  Baker:2006gm}. In this paper, we show that if the dark matter (DM)
consists of a scalar (fermion) of mass in the MeV range or lower, with
Yukawa couplings to neutrinos and another new fermion (scalar) with a
mass of a few MeV, the {\it en-route} resonant interaction of the DSNB
neutrinos with DM could also lead to a dip in the spectrum.  Such mass
range and  couplings are motivated by models in which neutrino masses
are generated radiatively~\cite{Boehm:2006mi, Farzan:2009ji,
  Farzan:2011tz}.  Although different theoretical models predict
slightly different shapes for the DSNB spectra, they are all smooth
spectra.  Thus, a sharp feature, such as a dip, would be a clear
signature of new physics~\cite{PalomaresWeiler, Weilertalk,
  PalomaresRuiztalk07, PalomaresRuiztalk11}.  Here we mainly discuss
the effects on the DSNB spectra, although we also mention the signature
from a galactic SN.

The paper is organized as follows. In Sec.~\ref{sec:nuinteractions},
we present the scenario and review  various bounds on its parameters.
In Sec.~\ref{sec:resonance}, we discuss the resonance absorption of
the DSNB and formulate the conditions for having a significant dip in
the spectrum. In Sec.~\ref{sec:spectra}, we discuss the evolution of
the flux, present the numerical results on the effect of the new
coupling on the DSNB spectrum and show the expected spectrum of events
at a detector such as HK.  In Sec.~\ref{sec:conclusions}, we review
our results and its implications for new physics and conclude.

\section{Neutrino interactions with Dark Matter}
\label{sec:nuinteractions}

Let us suppose that neutrinos have a coupling of form
\be{eq:coupling}
g N_R^\dagger \nu_L  \phi ~,
\ee
with a new scalar $\phi$ and a new fermion $N_R$, which are both
neutral and have a mass in the MeV range.  Of course, this coupling
can be only effective below the electroweak scale.  Various minimal
models to embed the coupling within a theory invariant under $SU(3)
\times SU(2) \times U(1)$ have been proposed~\cite{Ma:2006km,
  Farzan:2009ji, Farzan:2010mr}.  In addition, if the whole Lagrangian 
is invariant under a $Z_2$ symmetry ($N\to -N$, $\phi\to -\phi$ and
SM$\to$SM), the lightest of the particles $\phi$ and $N$ would be a DM
candidate.  The right-handed neutrino $N$ can be either of Dirac type
or of Majorana type.  The limiting case of pseudo-Dirac $N$, as
discussed below, is of particular interest.  The scalar particle
$\phi$ can be either real or complex.  In each case, two situations
are possible: $\mphi<\mn$ or $\mn<\mphi$.  Thus, there are in general
eight possibilities.

The SLIM scenario~\cite{Boehm:2006mi, Farzan:2009ji, Farzan:2011tz},
which links the neutrino mass with the DM relic density, corresponds
to the case that $\mphi<\mn$ with $\phi$ being real and $N$ Majorana.
In this scenario, the light neutrino masses are obtained at loop level
and the annihilation channels $\phi \phi \to \nu \nu,
\bar{\nu}\bar{\nu}$ determine the DM abundance.  Taking the observed
value of the DM abundance and neutrino masses in the range
$\sqrt{\Delta m_{\rm atm}^2}\sim 0.05~{\rm eV}-1~{\rm eV}$, it was
found that the mass of $N$ is in the $\sim$1--10~MeV
range~\cite{Boehm:2006mi, Farzan:2009ji, Farzan:2011tz}.  The mass of
$\phi$ would therefore be in the MeV range or lower.

There are a number of different observables that set bounds on the
coupling constants.  The coupling in Eq.~\rf{eq:coupling} can lead
to new decay modes for charged leptons and charged mesons such as
$K^+$ and $\pi^+$.  In particular, it can lead to $K^+ \to e^+ + N
+\phi$ and also $K^+ \to \mu^+ + N +\phi$, which would appear as decay
into charged leptons plus missing energy.  As long as ${\rm
  max}\{\mphi^2,\mn^2\} \ll m_{K, \pi}^2$, this discussion is similar
for all the eight possibilities enumerated above.  The present
bounds are $|g_e|^2<10^{-5}$~\cite{Ambrosino:2009aa} and
$|g_\mu|^2<10^{-4}$~\cite{Pang:1989ut, Farzan:2010wh}.  These bounds
could be improved by KLOE~\cite{KLOE} and NA62~\cite{NA62} data.  Of
course, these bounds
do not apply if the sum of the masses of $\phi$ and $N$ is larger than
the kaon mass.  For $m_K<m_\phi+m_N<m_D$, the strongest bound on $g_e$
comes from the $D$ meson decay modes: $|g_e|<0.4$~\cite{Lessa:2007up,
  Beringer:1900zz}.  From the $W$ boson decay
modes~\cite{Beringer:1900zz} we find $g_\mu,g_e <1$.  The bound on the
coupling of $\nu_\tau$ is much weaker as the $\tau$ leptons are not
produced in the kaon decays.  The strongest bound in this case comes
from $\tau$ decays~\cite{Lessa:2007up}.  In fact $g_\tau$ can be as
large as $O(1)$.

If $\phi$ is real and $N$ is of Majorana type, the active neutrino
mass would receive a contribution at one loop level. The upper bound
on masses of the active neutrinos then yields
$g<10^{-3}$~\cite{Boehm:2006mi, Farzan:2009ji, Farzan:2011tz}.  If
$\phi$ is complex or $N$ is of Dirac type, the lepton number would be
conserved so there would be no contribution to active neutrino masses.
A more interesting scenario is the case with pseudo-Dirac $N$ and
real $\phi$.  Let us restore flavor indices and assume a $U(1)\times
U(1)\times U(1)$ flavor symmetry softly broken only by $(m_R)_{\alpha
  \beta}$.  The flavor symmetry dictates that for any flavor
$\nu_\alpha$ there is a separate Dirac $N_\alpha$ with mass
$m_{N_\alpha} \bar{N}_\alpha N_\alpha$ and coupling $g_\alpha$.  The
flavor structure of the light active neutrinos would be given by
\be{eq:numass}
(m_\nu)_{\alpha \beta} \simeq \frac{g_\alpha g_\beta}{4\pi}
(m_R)_{\alpha \beta} \log (\frac{\Lambda^2}{m_{\phi, N}^2}) ~,
\ee
where $\Lambda$ is the cutoff scale above which the effective
coupling in Eq.~\rf{eq:coupling} is not valid (i.e., electroweak
scale) and $m_{\phi, N}^2$ is a linear combination of
$m_{N_\alpha}^2$, $m_{N_\beta}^2$ and $m_\phi^2$ of order of
$(1-10)^2$ MeV$^2$. By taking $m_R$ small enough, the bounds on the
coupling  from the mass of active neutrinos can be relaxed.  Notice
that even with $g_\mu\ll g_\tau$, we can obtain $(m_\nu)_{\mu \mu}\sim
(m_\nu)_{\tau \tau}$ provided that $(m_{R})_{\mu \mu}\gg (m_{R})_{\tau
  \tau}$.  Similar considerations hold for $g_e \ll g_\tau$.  Unless
stated otherwise, in the following, by $N$ we mean $N_\tau$, which has
the strongest coupling.

The coupling in Eq.~\rf{eq:coupling} also leads to the annihilation
of a DM pair.  If we require the DM production in the early universe
to be thermal, the total annihilation cross section should be
${\cal O}(1)$~pb.  As discussed in detail in the Appendix, for all the
cases discussed above this requirement implies $g\ll 0.1$, except when
$N$ is of (pseudo-)Dirac type and the DM candidate is a real scalar with
$m_\phi<m_N$.  Such a small coupling ($g\ll {\cal O}(0.1)$) would not
give a detectable dip in the DSNB (see below).  As a result, we focus
on the case with real $\phi$ and pseudo-Dirac $N$ in our analysis, for
which couplings ${\cal O}(0.1)$ are consistent with the thermal DM
scenario.  All in all, the overall behavior of our results for the
rest of the aforementioned cases is similar.

In addition, the cosmic microwave background data as well as the data
on big bang nucleosynthesis can set a lower bound on DM mass of the
order of MeV.  However, in the case of a real $ \phi$ as a DM
candidate these bounds are relaxed~\cite{Boehm:2013jpa}.  Stable
particles in the MeV range coupled to neutrino could also contribute
to SN cooling.  Within the present uncertainties in the SN models, a
real MeV mass scalar might be tolerated.

In summary, we will focus on the case with $N$ being of pseudo-Dirac
type and real $\phi$ playing the role of DM coupled dominantly to
$\nu_\tau$.  This setup escapes all the present bounds.  For
simplicity, with the purpose of clearly illustrating the effects under
discussion, in this work we only consider a single $\phi$ and a single
$N$.  In principle, in addition to active light neutrinos there can be
sterile neutrinos ($\nu_s$) coupled to $N$ and $\phi$: $g_s
N_R^\dagger \nu_s \phi$.  The $g_s$ coupling can be as large as ${\cal
  O}(1)$, increasing the decay width of $N$ dramatically.  We will
briefly discuss the implications of this case, too.

\section{Resonant absorption of supernova relic neutrinos}
\label{sec:resonance}

Following the discussion of the previous section, we will consider a
general coupling  of neutrinos and DM given by $g_i \nu_i \phi N$,
where $\nu_i$ represent mass eigenvalues of active neutrinos.  There
are two subtleties here: (1) As we saw in the previous section, for
$\phi$ and $N$ with masses in the MeV range, only the $\nu_\tau$
coupling can be significant, which means $g_i \propto U_{\tau i}$; (2)
as discussed in the previous section, reproducing the neutrino mass
structure requires more than one $N$.  Recalling that the couplings of
$N_e$ and $N_\mu$ are severely constrained by meson decay experiments,
their effects on the DSNB spectrum cannot be large.  We therefore
drop them from our discussion and focus on effects of a single $N$.

If the center-of-mass energy of the neutrino-DM system corresponds to
the $N$-pole ($\phi$-pole),  a resonant absorption along the
propagation of the DSNB would occur, giving rise to a dip in the
predicted flux. In this section, we formulate the conditions under
which the absorption dip can be significant.  Following the discussion
in the previous section, we assume that DM is composed of $\phi$.
Similar consideration holds for $\mn < \mphi$ with $N$ as the DM
candidate.  Thus, we use $\mr$ and $\mdm$ to refer to the mass of the
particle produced at resonance (in our case $\mn$) and the DM mass (in
our case $\mphi$), respectively.

The resonance neutrino energy in the laboratory frame, $\Eres$, is
\be{eq:Eres}
\Eres = \frac{\mr^2-\mdm^2}{2 \, \mdm} = E_0 \, (1+\zr) ~,
\ee
where $E_0$ is the energy of the relic neutrino observed at Earth and
$\zr$ is the redshift at which the resonant interaction occurred.  In
the above expression, we have neglected the very small momenta of the
DM particles due to nonzero temperature ($T/\mdm\ll 1$) and considered
them to be at rest.

The optical depth $\tau$ for the cosmic propagation of a relic SN
neutrino is given by
\be{eq:taudef}
\tau=\int \frac{c\,dt}{\lambda_\nu}
   =\int dz \,\frac{dt}{dz}\,n(z)\,\sigma (z)\,,
\ee
where $\lambda_\nu = (n \, \sigma)^{-1}$ is the mean-free path of the
relic SN neutrino, $n(z)$ is the DM density, $\sigma(z)$ is the
neutrino-DM interaction cross section and $dt/dz=-((1 + z) \,
H(z))^{-1}$ is the time-redshift relation.  For the epoch of interest,
\be{eq:Friedmann}
H(z) \simeq H_0 \, \sqrt{\Omega_\Lambda + \Omega_{\rm m,0} (1 + z)^3} ~.
\ee
with the present value of the Hubble parameter $H_0 = (67.3 \pm 1.2)
\times {\rm km~sec}^{-1}  {\rm Mpc}^{-1}$ and $\Omega_\Lambda =
0.685^{+0.018}_{-0.016}$ and $\Omega_{\rm m, 0} =
0.315^{+0.016}_{-0.018}$~\cite{Ade:2013zuv}.

The probability for a SN relic neutrino not to undergo interaction
during its cosmic propagation is given by $e^{-\tau}$, so the absorbed
fraction of flux, $f_{\rm abs}$, is given by $f_{\rm abs} =
1-e^{-\tau}$.  Hence, in order to determine the parameter range over
which significant dips in the DSNB flux could be produced, we should
impose some condition on $f_{\rm abs}$ (or equivalently on $\tau$).

The average DM density at any $z$ is given in terms of the present
density by
\be{eq:nz}
n(z) = n_0 \, (1 + z)^3 = \frac{\Omega_{\rm DM,0}\,\rho_c}{\mdm} \, (1
+ z)^3 \simeq 1.26 \, \left( \frac{{\rm keV}}{\mdm} \right) \, (1 +
z)^3 \, {\rm cm}^{-3}
\ee
where $\rho_c = 3H_0^2/(8\,\pi\,G_N) = 4.77 \, {\rm keV/cm}^3$ is the
critical density of the Universe and $\Omega_{\rm DM,0} = 0.265$ is the
present fraction of DM~\cite{Ade:2013zuv}.

Since the non-resonant processes have negligible effects for the
DSNB absorption\footnote{The non-resonant part can be estimated as
  $\sigma_{\rm nr}\sim {g^4}/(16 \pi E_\nu^2)$ so $\tau_{\rm nr}\sim
  \int dz({dt}/{dz}) n(z) \sigma_{\rm nr}(z)\sim 4 \, g^4$, which is
  negligible for $g<0.5$.}, we shall only consider the s-channel
contribution to the cross section.  For the case when the intermediate
resonance particle decays isotropically in its rest frame (e.g., if it
is a scalar), the decay of the resonance produces a flat final
neutrino spectrum over the interval $E_{min} \le E'_\nu \le E_\nu$.
We can therefore write the differential cross section as
\be{eq:dsdeflat}
\frac{d\sigma^p_{i j}}{dE_\nu'} (E_\nu, E_\nu') = \sigma^p_{i
  j}(E_\nu) \, \frac{\theta (E_\nu - E_\nu')}{E_\nu-E_{\rm min}}
\theta(E_\nu'-E_{\rm min}) ~,
\ee
where the superindex $p=\rm{LC, LV}$ refers to the lepton number
conserving (LC) or lepton number violating (LV) scattering, $E_{\rm
  min} = \mdm E_\nu/(2E_\nu+\mdm)$ and $\theta$ is the step function.
The $i$ ($j$) subindex refers to the incoming (outgoing) neutrino.
However, in our case $N$ (being a fermion) decays non-isotropically.
In general, one can write
\be{eq:dsde}
\frac{d\sigma^p_{i j}}{dE^\prime} = \frac{d\sigma^p_{i j}}{d\cos\theta}
\frac{2E_\nu+\mdm}{E_\nu^2}
\ee
where $d\sigma^p_{i j}/d\cos\theta$ is the partial scattering cross
section in the center of mass frame of $\phi$-$N$ system and
\be{eq:erange}
\frac{\mdm}{2 E_\nu+\mdm}E_\nu<E_\nu^\prime <E_\nu  \;.
\ee
Close to the resonance energy, the cross section of $\nu_i \phi \to N
\to \nu_j \phi$ is given by
\be{eq:slc}
\frac{d \sigma^{\rm LC}_{i j}}{d\cos\theta} = \frac{g_{i}^2 g_{j}^2}{32
  \pi} \, \frac{ (\mr^2-\mdm^2)^2}{(\mr^2 + \mdm^2)} \,
\frac{1+\cos\theta}{(s-\mr^2)^2+\gr^2 \, \mr^2} ~,
\ee
where $\gr$ is the decay width of the particle produced at resonance
(in our case $N$) and $s$ is the Mandelstam variable, $s = 2 \mdm
E_\nu + \mdm^2$ in which $E_\nu$ is the energy of $\nu$ at the
interaction.

If $N$ is of Majorana type, in addition to LC scattering $\nu_i \phi
\to N \to \nu_j \phi$, we can have $\nu_i \phi \to N \to \bar{\nu}_j
\phi$ as well as $\bar{\nu}_i \phi \to N \to \nu_j \phi$ with
\be{eq:slv}
\frac{d \sigma^{\rm LV}_{i j}}{d\cos\theta} = \frac{g_{i}^2 g_{j}^2}{32
  \pi} \, \frac{ (\mr^2-\mdm^2)^2}{(\mr^2+\mdm^2)} \,
\frac{1-\cos\theta}{(s-\mr^2)^2+\gr^2 \, \mr^2} ~.
\ee
Of course, for Dirac $N$ and complex $\phi$, lepton number is
conserved and the channel corresponding to Eq.~\rf{eq:slv} is
forbidden.  As discussed in the Appendix, in the pseudo-Dirac case,
practically for each pair $(N_R~N_L)$, there are two quasi-degenerate
mass eigenstates $N_1$ and $N_2$ with couplings $g/\sqrt{2}$ and
$ig/\sqrt{2}$, so their contributions to the LV process approximately
cancel each other.  Thus, the LV processes for the pseudo-Dirac
scenario can be neglected.  Having two mass eigenvectors, there should
be, in principle, two resonances.  However if the mass difference is
much smaller than $\gr$, the two peaks cannot be resolved and the LC
cross section is given mainly by Eq.~\rf{eq:slc}.  If the active
neutrino masses are produced by couplings as those in
Eq.~\rf{eq:coupling}, from Eq.~\rf{eq:numass} we conclude
$|m_{N_1}-m_{N_2}|/\gr=0.06 (m_\nu/0.05~{\rm eV})({\rm
  MeV}/m_N)(0.1/g_i)^4\ll 1$, so we are in the limit that we can
safely use Eq.~\rf{eq:slc}.

Close to the resonance, the total LC cross section is given by
\be{eq:total}
\sigma_{i j} (s) \simeq \frac{g_i^2g_j^{2}}{16 \pi}
\frac{(\mr^2-\mdm^2)^2}{\mr^2+\mdm^2} \frac{1}{(s-\mr^2)^2+\gr^2 \,
  \mr^2} ~.
\ee
Notice that for Majorana $N$, the cross section is twice as much the
one in Eq.~\rf{eq:total}.  If the main decay mode of $N$ is to $\phi
\nu$, we obtain
\be{eq:GammaR}
\gr   = \sum_i \frac{g_i^2}{16\pi}\frac{(\mr^2-\mdm^2)^2}{\mr^3} ~.
\ee
For $\gr \ll \mr$, it is convenient to use the narrow width
approximation limit to analytically solve Eq.~\rf{eq:taudef}.  Thus,
the cross section can be written as
\ba{eq:totalNWA}
\sigma_{i j} (s) & \simeq & \frac{g_i^2g_j^{2}}{\sum_k g_k^2} \,
\pi \, \frac{\mr^2}{\mr^2+\mdm^2}\delta\left(s-\mr^2\right) \nonumber \\
& = & \frac{g_i^2g_j^{2}}{\sum_k g_k^2} \, \pi \,
\frac{1+z}{\mr^2-\mdm^2} \, \frac{\mr^2}{\mr^2+\mdm^2}
\ \delta\left((1+z) - \frac{\mr^2-\mdm^2}{2 \mdm E_0}\right) ~.
\ea

Inserting Eqs.~\rf{eq:Friedmann},~\rf{eq:nz} and~\rf{eq:totalNWA} into
Eq.~\rf{eq:taudef}, the optical depth for a given neutrino mass
eigenstate emitted at a redshift of $\zr$ with an energy of $E_0
(1+\zr)$ can be analytically calculated.  For $E_0 \le \Eres$, we have
\ba{eq:tau}
\tau_i(\zr) & = & \sum_j \, \frac{g_i^2g_j^{2}}{\sum_k
g_k^2} \, \left(\frac{\pi}{\mr^2-\mdm^2}\right) \,
\left(\frac{\mr^2}{\mr^2+\mdm^2} \right) \,
\left(\frac{n_0}{H_0}\right) \, \left(\frac{\Omega_{\rm
    DM}(\zr)}{\Omega_{\rm DM,0}}\right) \nonumber \\
& \simeq & 5 \times 10^2 \,  g_i^2 \, \left( \frac{20~{\rm
MeV}}{\Eres}\right) \,  \left(\frac{\rm MeV}{\mdm}\right)^2
\left(\frac{\Eres + \mdm/2}{\Eres + \mdm}\right)\,
\left(\frac{\Omega_{\rm DM}(\zr)}{\Omega_{\rm DM,0}}\right) ~,
\ea
where $\Omega_{\rm DM}(z) = \Omega_{\rm DM,0}
(1+z)^3/\sqrt{\Omega_\Lambda + \Omega_{\rm m,0} (1+z)^3}$.  Notice
that the fraction of DM, $\Omega_{\rm DM}(z)$, is a monotonously
increasing function with $\Omega_{\rm DM}(0) = \Omega_{\rm DM,0}$ and
$\Omega_{\rm DM}(6) = 32.9 \, \Omega_{\rm DM,0}$.

For a minimal absorption of $f_{\rm abs}=10\%$, we obtain the condition
\be{eq:fom}
g_i^2 > 2 \times 10^{-4} \left(\frac{\Eres}{20 ~{\rm MeV}}\right) \,
\left(\frac{\mdm}{\rm MeV}\right)^2 \left(\frac{\Omega_{\rm
    DM,0}}{\Omega_{\rm DM}(\zr)}\right) \, \left(\frac{\Eres +
  \mdm}{\Eres + \mdm/2}\right) ~.
\ee

\begin{figure}[t]
\begin{center}
\includegraphics[width=0.8\linewidth]{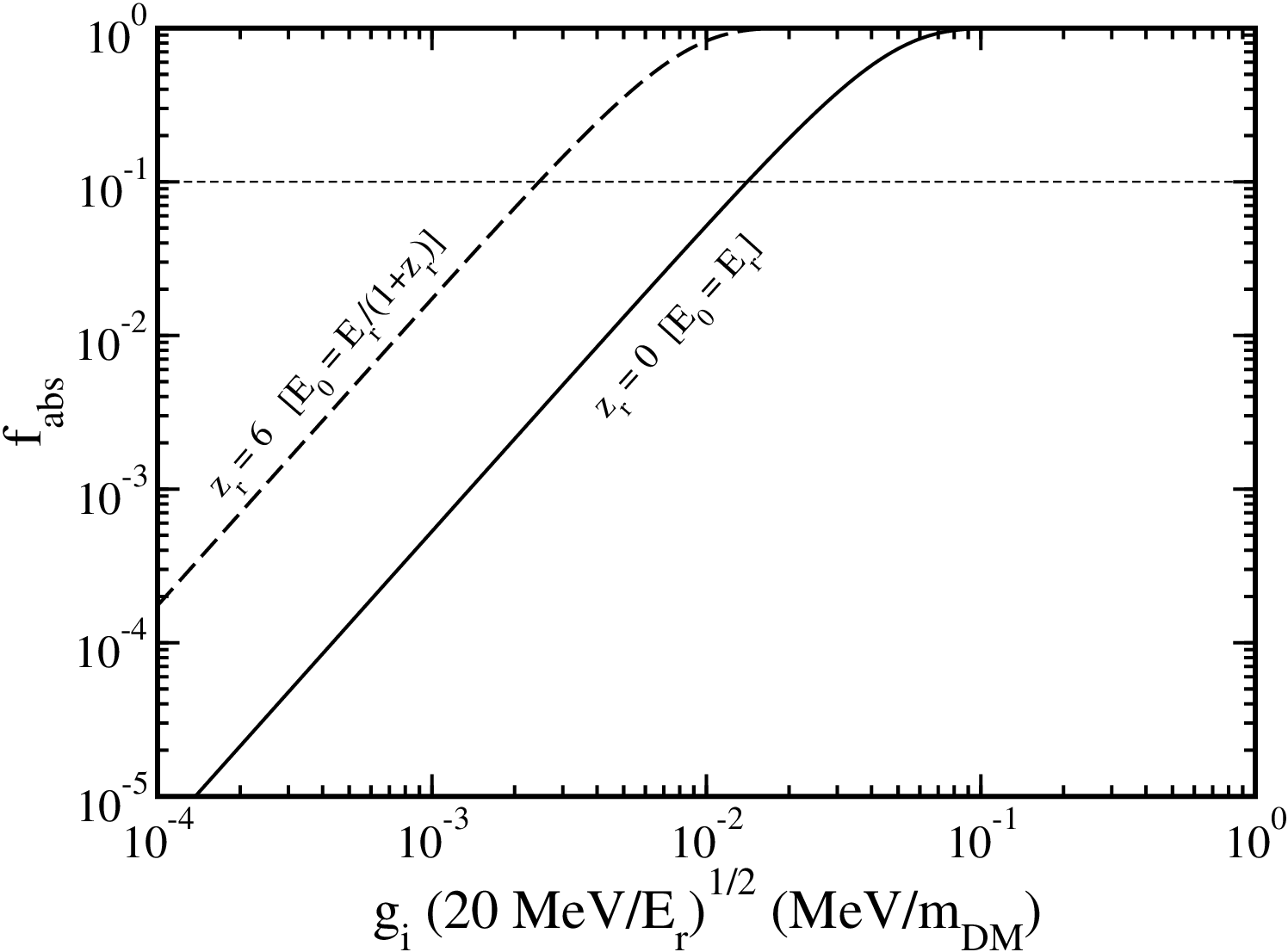}
\caption{\sl The absorbed fraction of the flux at the resonance energy,
  $f_{\rm abs} (\zr)= 1 -e^{-\tau_i (\zr)}$, as a function of the coupling
for two different redshifts. The thin dashed line indicates 10\%
absorption. Here we assume $\Eres \gg \mdm$.}
\label{fig:fabs}
\end{center}
\end{figure}

In Fig.~\ref{fig:fabs}, assuming $\Eres \gg \mdm$, we show the absorbed
fraction of the flux at the resonance energy as a function of the
coupling for two redshifts.  As seen from the figure, while for $g_i
\lesssim 10^{-2} \, (\Eres/20 \, {\rm MeV})^{1/2} (\mdm/{\rm MeV})$,
the absorption at $z=0$ is negligible, for slightly larger values of
the coupling, $g_i \gtrsim {\rm few} \times 10^{-2} \, (\Eres/20 \, {\rm
  MeV})^{1/2} (\mdm/{\rm MeV})$, the absorption becomes significant.
Increasing the redshift, absorption for a given coupling becomes
stronger, but occurs at a lower observed energy.  The sharp dependence
on the coupling reflects the exponential dependence of $f_{\rm abs}$
on $g_i^2$.

So far we have considered the cumulative absorption of the DSNB.
Nevertheless, the DM density in halos is much larger than the average
DM density in the Universe, so one could wonder whether neutrinos
produced by a SN in our own galaxy, traveling towards us in a medium
with high DM density, will experience similar effects.  In fact, the
optical depth for neutrinos in the Milky Way could be much larger than
the one corresponding to cosmological distances with average density
given by Eq.~\rf{eq:nz}.  However, only neutrinos with energies in a
very narrow interval, ($\Eres-\gr \, \mr/2\mdm, \, \Eres+\gr \,
\mr/2\mdm$), would undergo resonant scattering. Thus, as long as $\gr
\, \mr/2\mdm \ll O(1)$~MeV, one can neglect the effect of resonant
neutrino absorption in the DM halo of the Milky Way because the dip
would be smeared due to the finite energy resolution of the detector.
Let us note however that, if the resonantly produced particle ($N$ in
this case) has other decay modes which increase $\gr$ to $O(1)$MeV,
this effect might be detectable.  One example can be $N$ decaying to
$\phi$ and a sterile neutrinos $\nu_s$ with $\Gamma(N\to \nu_s
\phi)\sim 1$~MeV.

In a similar way, SN explosions at cosmological distances have also
taken place inside galaxies, so right after their production,
neutrinos traverse a medium with a much higher DM density than the
average cosmic background. Thus, large resonant absorption at the host
galaxy at redshift $z$ is expected to occur in the observed energy
interval ($\Eres/(1+z)-\gr/(1+z) \, \mr/2\mdm, \, \Eres/(1+z) +
\gr/(1+z) \, \mr/2\mdm$).  However, for a very narrow resonance, the
fraction of the flux which is absorbed at the host is so small that it
amounts to a very small suppression of the cumulated flux.  This is
unlike the case of the redshift-integrated effect in the DSNB for
which all energies between $\Eres$ and $\Eres/(1+z)$ experience
resonant absorption at some point on their way from the host at $z$ to
us.

\section{Neutrino spectra and events}
\label{sec:spectra}

Let us consider the process in which a neutrino of mass $m_i$, with
energy ${\cal E}_z = E_\nu (1+z)$ hits a DM particle $\phi$ and
resonantly produces $N$ at redshift $z$.  Subsequently, in the case of
lepton number conserving (violating) processes, $N$ decays into a
neutrino (antineutrino) of mass $m_j$, with energy ${\cal E}_z' =
E_\nu' (1+z)$ and a DM particle $\phi$, i.e., $\nu_i \phi \to N \to
\nu_j \phi$ ($\nu_i \phi \to N \to \bar\nu_j \phi$).  Taking $\Phi_i$
($\Phi_{\bar i}$) as the neutrino (antineutrino) flux of mass $m_i$,
let us define
\be{eq:defs}
F_i (t, E_\nu)  \equiv \frac{d\Phi_i}{dE_\nu} (t, E_\nu) ~.
\ee
The time evolution of $F_i (t, E_\nu) $ is then governed by (see,
e.g., Ref.~\cite{Berezinsky:2005fa})
\ba{eq:main}
\frac{\partial F_i (t, E_\nu)}{\partial t}  & = & -  3
  H(t) F_i (t, E_\nu)  +
\frac{\partial}{\partial E_\nu} \left(H(t) E_\nu \, F_i (t, E_\nu)
\right) - \frac{1}{\lambda_i (t, E_\nu)} F_i (t, E_\nu)
\nonumber \\
 & &  + \sum_j \int_{E_\nu}^{\infty} \, dE_\nu' \, \Big[ {\cal T}_{j
  i}^{\rm LC} (t, E_\nu', E_\nu)  \, F_j(t, E_\nu') + {\cal T}_{j
  i}^{\rm LV} (t, E_\nu', E_\nu)  \, F_{\bar j} (t, E_\nu') \Big]
\nonumber \\
& & + {\cal L}_i (t, E_\nu)/a^3 (t) \, ,
\ea
where ${\cal L}_i (t, E_\nu)$ in the last term is the comoving
luminosity of the source of neutrinos of mass $m_i$ and
\ba{eq:lambda}
\lambda_i (t, E_\nu) & \equiv & \frac{1}{\sum_{p,j} \, n(t)
  \, \sigma^p_{i j} (E_\nu)} \hspace{5mm} ; \, p={\rm LC, LV} \\
\label{TLC} {\cal T}_{j i}^{\rm LC} (t, E_\nu', E_\nu) & \equiv &
n(t) \, \frac{d\sigma^{\rm LC}_{j i}}{dE_\nu} (E_\nu', E_\nu)   \\
\label{TLV} {\cal T}_{j i}^{\rm LV} (t, E_\nu', E_\nu) & \equiv &
n(t) \, \frac{d\sigma^{\rm LV}_{\bar j i}}{dE_\nu} (E_\nu', E_\nu)\ .
\ea

In the right hand side of Eq.~\rf{eq:main}, the first and second terms
originate from the expansion of the Universe and the induced adiabatic
energy losses, respectively.  The third term represents the absorption
dip, where $\lambda_i (t, E_\nu)$ is the mean-free path of neutrinos
of mass $m_i$ given in Eq.~\rf{eq:lambda}.  In Eq.~\rf{eq:lambda},
$n(t)$ is the number density of the DM particles at time $t$,
Eq.~\rf{eq:nz}, and $\sigma^p_{i j}$ is the total cross section of the
considered process ($p$=LC, LV).  The fourth term in Eq.~\rf{eq:main}
represents the repopulation of the spectrum at energies lower than the
resonance energy, where the contributions from both LC and LV
processes are included.  Obviously, for observed energies at the Earth
larger than the resonance energy, the original spectrum does not
suffer distortion, but just becomes redshifted.  Finally, the fifth
term in Eq.~\rf{eq:main} represents the luminosity of the sources.

The comoving luminosity of the source of neutrinos of flavor $\alpha$
at redshift $z$, ${\cal L}_\alpha (z, E_\nu)$, is given by
\be{eq:luminosity}
{\cal L}_\alpha (z, E_\nu) = R_{\rm SN} (z) \, F_\alpha^{\rm SN} (E_\nu)
\ee
where  $F_\alpha^{\rm SN} (E_\nu)$ is the number spectrum of neutrinos
of flavor $\alpha$ emitted by a typical SN and $R_{\rm SN} (z)$
represents the SN rate per comoving volume at redshift $z$.

For the SN rate per comoving volume we invoke canonical parameters for
optically luminous core-collapse SN ($M_{\rm min} = 8 M_\odot$ and
$M_{\rm max} = 40 M_\odot$)~\cite{Horiuchi:2011zz} and use the fit to
the star formation rate from the combination of low-$z$ ultraviolet
and far infrared data~\cite{Hopkins:2006bw} and from high-$z$ galaxies
and gamma-ray bursts data, assuming a Salpeter initial mass function,
obtained by Ref.~\cite{Kistler:2013jza},
\be{eq:SNrate}
R_{\rm SN} (z) = 0.0088 \, M_\odot^{-1} \, \, \dot{\rho}_0 \,
\left[(1+z)^{a \, \zeta} + \left(\frac{1+z}{B}\right)^{b \, \zeta} +
  \left( \frac{1+z}{C}\right)^{c \, \zeta} \right]^{1/\zeta} ~,
\ee
with $\dot{\rho}_0 = 0.02 \, M_\odot \, {\rm yr}^{-1} \, {\rm
  Mpc}^{-3}$, $a=3.4$, $b=-0.3$, $c=-2.5$, $\zeta=-10$,
$B=(1+z_1)^{1-a/b}$ and $C=(1+z_1)^{(b-a)/c} \, (1+z_2)^{1-b/c}$ in
which $z_1=1$ and $z_2=4$.

For the neutrino spectrum from a typical SN, we consider the
parameterization for each flavor given by~\cite{Keil:2002in}
\be{eq:SNispec}
F_\alpha^{\rm SN} (E_\nu) =
\frac{(1+\beta_{\nu_\alpha})^{1+\beta_{\nu_\alpha}} \,
  L_{\nu_\alpha}}{\Gamma(1+\beta_{\nu_\alpha}) \,
  \overline{E}_{\nu_\alpha}^2} \,
\left(\frac{E_\nu}{\overline{E}_{\nu_\alpha}}\right)^{\beta_{\nu_\alpha}}
\, e^{-(1+\beta_{\nu_\alpha}) E_\nu/\overline{E}_{\nu_\alpha}} ~.
\ee
Below, we show results for two sets of parameters.  We take the
optimistic case from the simulation of the Lawrence Livermore
group~\cite{Totani:1997vj} (model A) with relatively high average
energies: $\overline{E}_{ \nu_e} = 11.2$~MeV, $\overline{E}_{\bar
\nu_e} = 15.4$~MeV, $\overline{E}_{\nu_x} = 21.6$~MeV; $\beta_{\nu_e}
= 2.8$, $\beta_{\bar \nu_e} = 3.8$, $\beta_{\nu_x} = 1.8$; $L_{\nu_e}
=  L_{\bar \nu_e} = L_{\nu_x} =
5.0\times10^{52}$~ergs~\cite{Ando:2004sb}, in which $\nu_x$ represents
non-electron-flavor neutrinos and antineutrinos.  Notice however that
this simulation overlooks some relevant neutrino processes. Recent
simulations indicate lower average energies for the $\bar \nu_e$ and
$\nu_x$ flavors.  Hence, we also study model B with $\overline{E}_{
  \nu_e} = 10$~MeV, $\overline{E}_{\bar \nu_e} = 12$~MeV,
$\overline{E}_{\nu_x} = 15$~MeV; $\beta_{\nu_e} = 3$, $\beta_{\bar
  \nu_e} = 3$, $\beta_{\nu_x} = 2.4$; $L_{\nu_e} = L_{\bar \nu_e} =
L_{\nu_x} = 5.0\times10^{52}$~ergs~\cite{Lunardini:2012ne,
  Tamborra:2012ac}.  The parameters are summarized in
Tab.~\ref{tab:tab1}.

\begin{table}[t]
\begin{center}
\begin{tabular}{|c|c|c|c|c|c|c|}
\hline ~ & $\overline{E}_{ \nu_e}$ [MeV] & $\overline{E}_{\bar \nu_e}$
       [MeV] & $\overline{E}_{\nu_x}$ [MeV] & \, $\beta_{\nu_e}$ \, &
       \, $\beta_{\bar \nu_e}$ \, & \, $\beta_{\nu_x}$ \, \cr
\hline \hline Model A~\cite{Totani:1997vj} & 11.2 & 15.4 & 21.6 & 2.8
& 3.8 & 1.8
\cr Model B~\cite{Lunardini:2012ne, Tamborra:2012ac} & 10 & 12 & 15 &
3 & 3 & 2.4 \cr \hline
\end{tabular}
\caption{Parameters for the different neutrino spectra according to the
  parameterization given in Eq.~\rf{eq:SNispec}, for two different
  models.  The luminosity of the different flavors for both models are
  taken to be $L_{\nu_e} = L_{\bar \nu_e} = L_{\nu_x} =
  5.0\times10^{52}$~ergs.}
\label{tab:tab1}
\end{center}
\end{table}

Neutrino fluxes arise from the central regions of the collapsed star
where the density is very high and  the effective neutrino mixings
are therefore strongly  suppressed.  For the measured values of the
neutrino mixing parameters, the propagation inside the star is
adiabatic. Thus,  at the surface of the star, the fluxes of mass
eigenstates can be identified with the flavor fluxes at
production\footnote{Let us note that neutrino-neutrino collective
  effects introduce corrections below the 10\% level on the DSNB, with
  no energy-dependent signatures due to smearing over time and over
  the SN population~\cite{Lunardini:2012ne}.  In this work, we neglect
  this small correction.}.  In the case of normal hierarchy (NH) for the
neutrino mass ordering, $\overline{\nu}_e$ ($\nu_e$) is coincident
with $\overline{\nu}_1$ ($\nu_3$), whereas in the case of inverted
hierarchy (IH), $\overline{\nu}_e$ ($\nu_e$) coincides with
$\overline{\nu}_3$ ($\nu_2$), so in terms of the neutrinos with
definite mass, the SN spectra are~\cite{Dighe:1999bi}
\ba{eq:spectrummassNH}
F_{\bar \nu_1}^{\rm SN} (E_\nu) = F_{\bar \nu_e}^{\rm SN} \, ; \, \, &
F_{\bar \nu_2}^{\rm SN} (E_\nu) = F_{\nu_x}^{\rm SN} \, ; \, \, &
F_{\bar \nu_3}^{\rm SN} (E_\nu) = F_{\nu_x}^{\rm SN} \nonumber \\
F_{\nu_1}^{\rm SN} (E_\nu) = F_{\nu_x}^{\rm SN} \, ; \, \, &
F_{\nu_2}^{\rm SN} (E_\nu) = F_{\nu_x}^{\rm SN} \, ; \, \, &
F_{\nu_3}^{\rm SN} (E_\nu) = F_{\nu_e}^{\rm SN} ~,
\ea
for NH, and
\ba{eq:spectrummassIH}
F_{\bar \nu_1}^{\rm SN} (E_\nu) = F_{\nu_x}^{\rm SN} \, ; \, \, &
F_{\bar \nu_2}^{\rm SN} (E_\nu) = F_{\nu_x}^{\rm SN} \, ; \, \, &
F_{\bar \nu_3}^{\rm SN} (E_\nu) = F_{\bar \nu_e}^{\rm SN} \nonumber \\
F_{\nu_1}^{\rm SN} (E_\nu) = F_{\nu_x}^{\rm SN} \, ; \, \, &
F_{\nu_2}^{\rm SN} (E_\nu) = F_{\nu_e}^{\rm SN} \, ; \, \, &
F_{\nu_3}^{\rm SN} (E_\nu) = F_{\nu_x}^{\rm SN} ~,
\ea
for IH.

The solution to Eq.~\rf{eq:main} for neutrinos (and equivalently for
antineutrinos) in terms of neutrino mass eigenstates is given by (see,
e.g., Ref.~\cite{Ahlers:2009rf}):
\ba{eq:solution}
F_i (z, E_\nu) & = & (1+z)^2 \, \int_z^{z_{\rm max}} \frac{dz'}{H(z')}
e^{-\int_z^{z'}
  \frac{dz''}{(1+z'') H(z'')} \frac{1}{\lambda_i(z'', {\cal E}_{z''})}}
\times \bigg\{ {\cal L}_i (z', {\cal E}_{z'}) \nonumber \\
& & + \sum_j \int_{{\cal E}_{z'}}^{\infty} \, d{\cal E}_{z'}'
  \Big[ \,{\cal T}_{j i}^{\rm LC} (z', {\cal E}_{z'}', {\cal E}_{z'}) \,
  F_j (z', {\cal E}_{z'}') + {\cal T}_{j i}^{\rm LV} (z', {\cal
    E}_{z'}', {\cal E}_{z'}) \, F_{\bar j} (z', {\cal E}_{z'}')
  \Big] \bigg\} ~,
\ea
where we take $z_{\rm max} = 6$.  Note that most of the signal comes
from SN explosions at $z<1$, so the exact upper limit for the redshift
is not crucial.

Finally, the spectrum of neutrinos of flavor $\alpha$ at Earth is
given by
\be{eq:spectrumflavor}
F_\alpha (z=0, E_\nu) = \sum_i |U_{\alpha i}|^2 \, F_i (z=0, E_\nu) ~.
\ee
where $U$ is the PMNS neutrino mixing matrix. We use
$\sin^2\theta_{12} = 0.306$, $\sin^2\theta_{23} = 0.446$ ($0.587$),
$\sin^2\theta_{13} = 0.0229$ and $\delta=0$ for NH
(IH)~\cite{GonzalezGarcia:2012sz} (see also
Refs.~\cite{Tortola:2012te, Fogli:2012ua}).

\begin{figure}[t]
\includegraphics[width=0.47\linewidth]{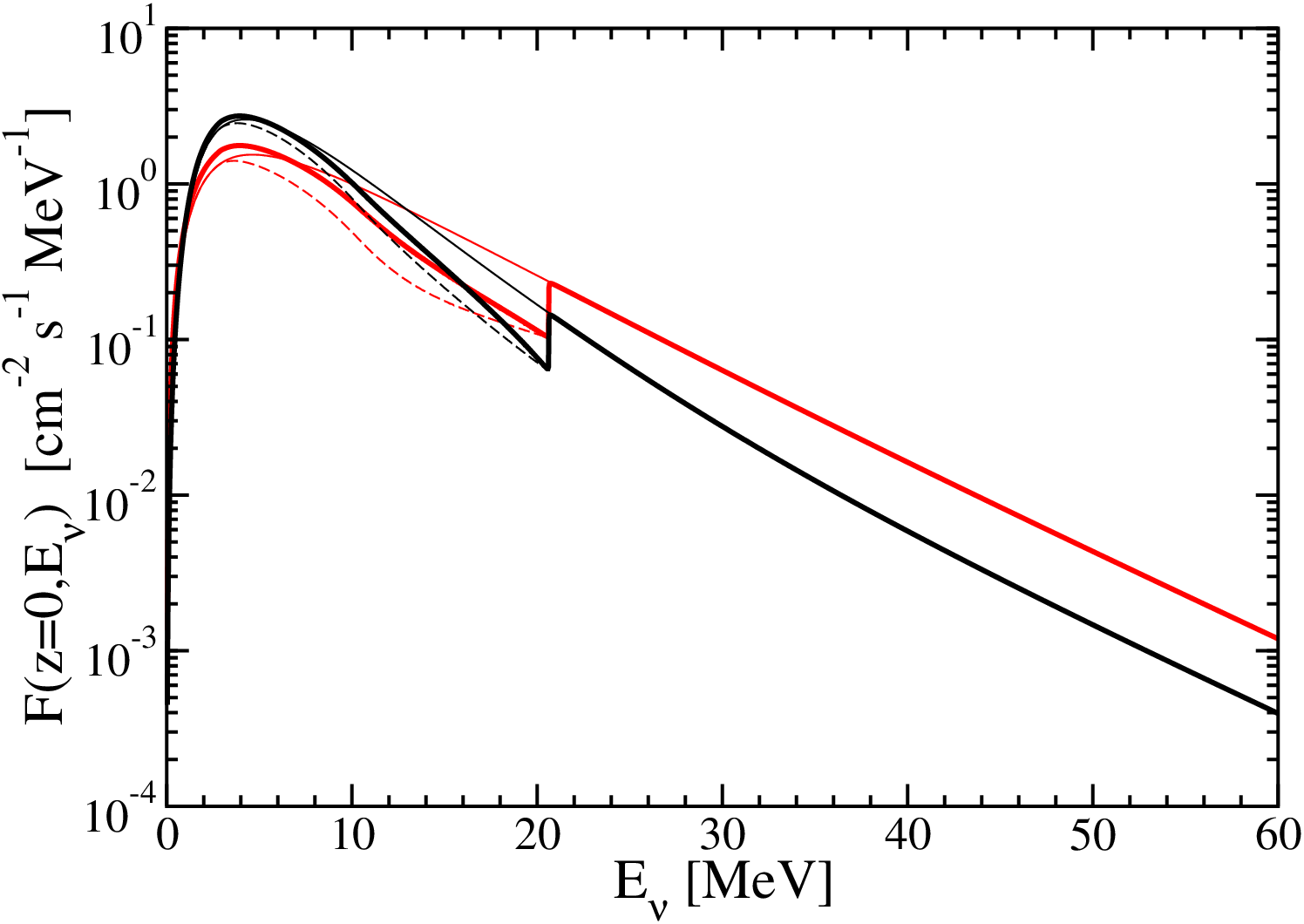}
\hspace{5mm}
\includegraphics[width=0.47\linewidth]{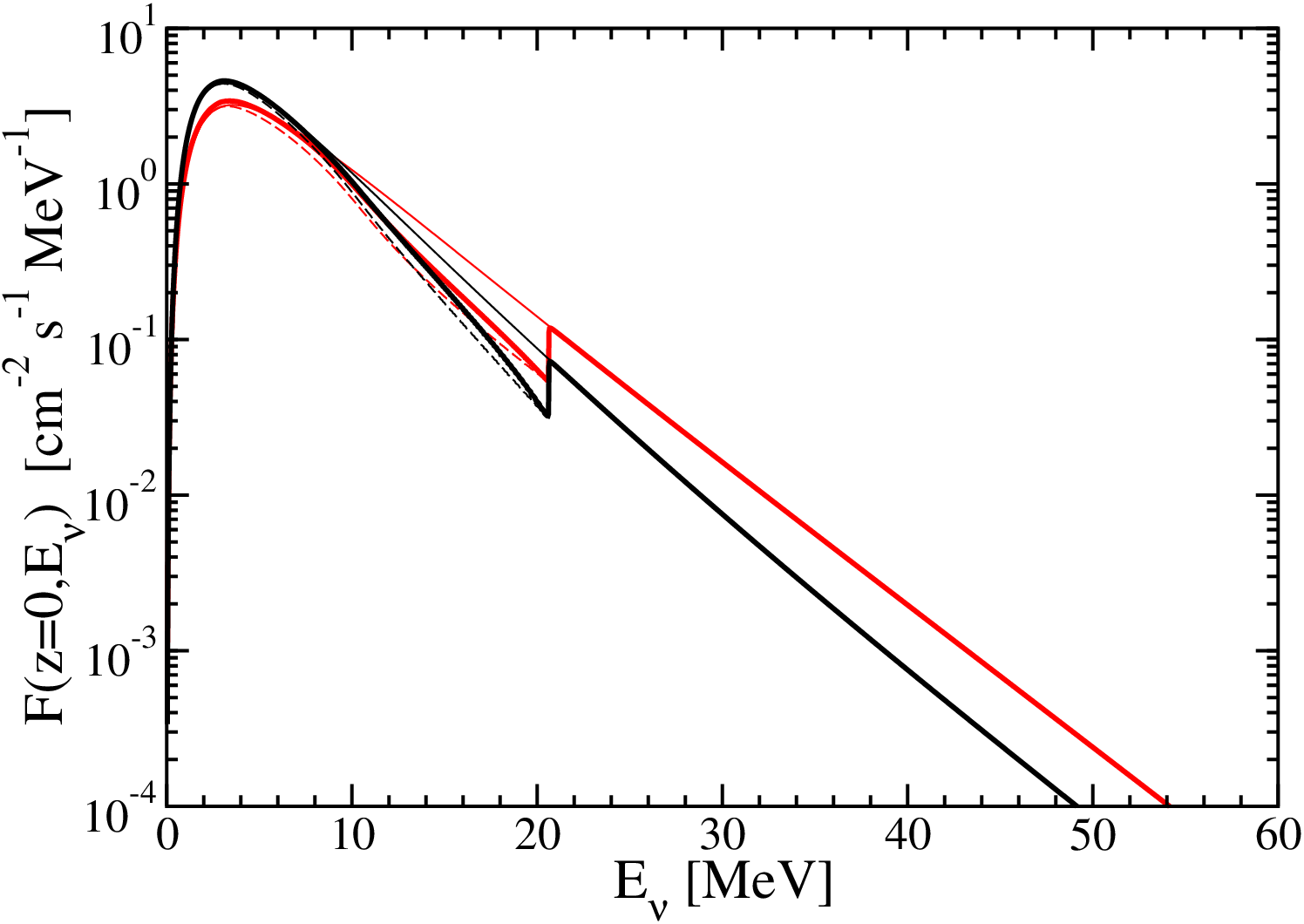} \\
\vspace{9mm}
\includegraphics[width=0.47\linewidth]{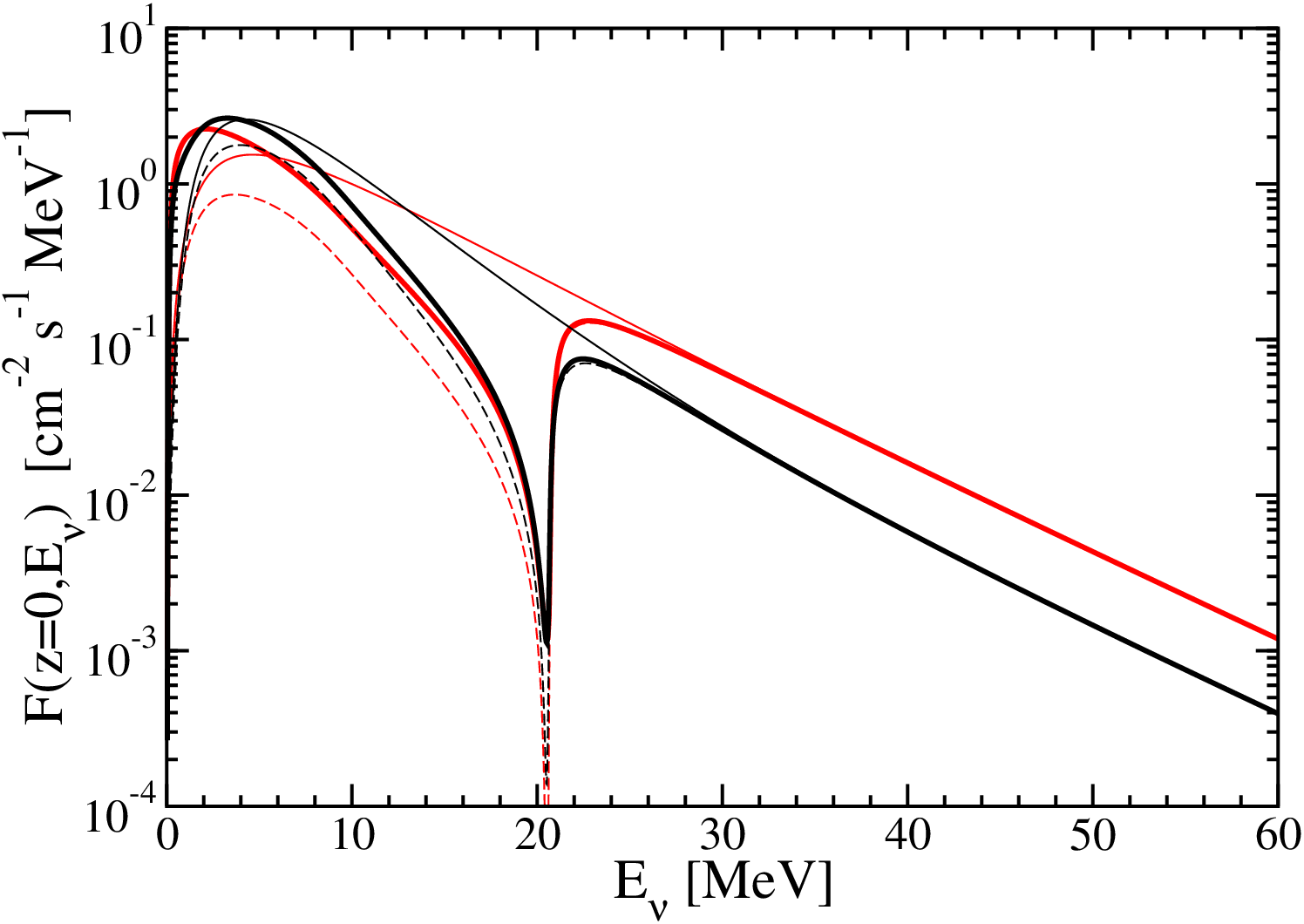}
\hspace{5mm}
\includegraphics[width=0.47\linewidth]{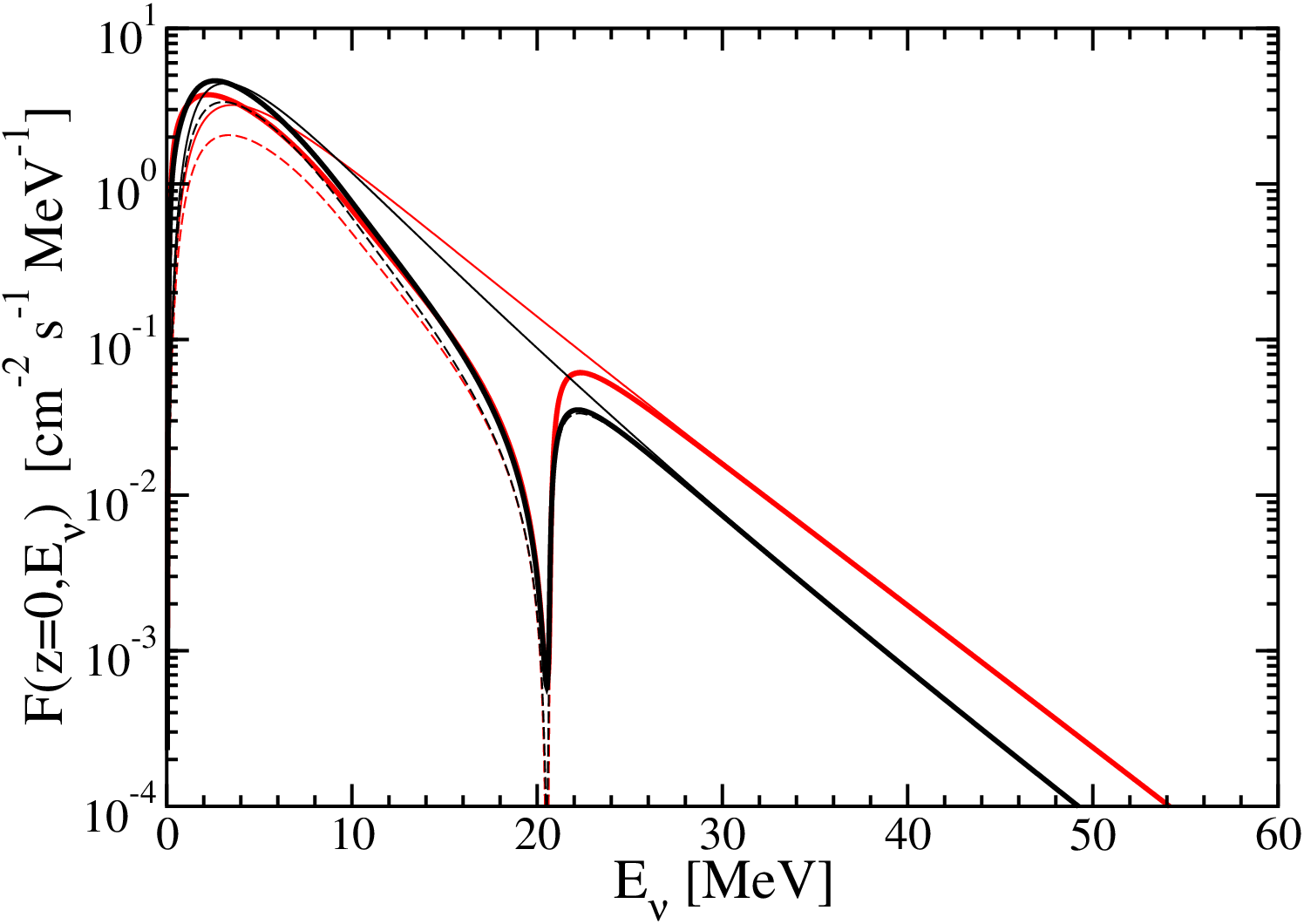}
\caption{{\sl DSNB $\bar{\nu}_e$ spectra for model A (left panels) and
    model B (right panels), assuming $g_\tau = 0.1$ (upper panels) and
    $g_\tau = 0.5$ (lower panels).  In the four panels, we assume
    $\mphi = \mdm=1$~MeV, $\mn = \mr=6.5$~MeV (Dirac) and $g_e = g_\mu
    = 0$, for NH (lower black lines at high energies) and IH (upper
    red lines at high energies).  The effect of the
    redshift-integrated absorption is depicted by the thick solid
    lines.  The case of no absorption is represented by thin solid
    lines, whereas dashed lines do not include the redistribution of
    the flux to lower energies after the interaction (e.g.,
    $\bar{\nu}_i \phi \to N \to \bar{\nu}_s \phi$).}}
\label{fig:spectra}
\end{figure}

In Fig.~\ref{fig:spectra} we show the resulting integrated
$\bar{\nu}_e$ spectra for model A (left panels) and model B (right
panels) for the two neutrino mass hierarchies\footnote{For the
  numerical calculations we use Eqs.~\rf{eq:slc}-\rf{eq:total}.}.  We
show the spectra after cosmological absorption and redistribution to
lower energies (thick solid lines), with no absorption considered (thin
solid lines) and with absorption, but without flux redistribution
(dashed lines). The absence of flux redistribution takes place e.g.,
when the decay of $N$ is dominated by a coupling to sterile neutrinos:
$\bar{\nu}_i \phi \to N \to \bar{\nu}_s \phi$.  We consider $N$ to be
a Dirac particle and assume $\mphi = \mdm = 1$~MeV, $\mn = \mr =
6.5$~MeV ($\Eres = 20.6$~MeV) and $g_e = g_\mu = 0$, $g_\tau = 0.1$
(upper panels) and $g_\tau = 0.5$ (lower panels).  In the mass basis,
the couplings read $g_1=g_\tau U_{\tau 1}\simeq 0.03$, $g_2=g_\tau
U_{\tau 2}\simeq 0.06$ and  $g_3=g_\tau U_{\tau 3}\simeq 0.07$.
Notice that although $g_e$ is taken to be zero, the spectrum of
$\bar{\nu}_e$ is affected due to neutrino mixing.  The figures show
that, as expected, the distortion of the spectrum starts at the
resonance energy.  The dip is broadened due to the absorption at
different redshifts.  Above $\sim$10~MeV the flux is higher for the
case of IH.  This can be easily understood from
Eqs.~\rf{eq:spectrummassNH}, \rf{eq:spectrummassIH} and
\rf{eq:spectrumflavor}. In the case of IH, the observed $\bar{\nu}_e$
flux is dominated by the $\nu_x$ component ($|U_{e3}|^2 \simeq 0.02$),
which has the highest average temperature, whereas in the case of NH,
$\sim $70\% ($|U_{e1}|^2 \simeq 0.7$) of the original $\bar{\nu}_e$
flux survives.  Thus, the final $\bar{\nu}_e$ flux at Earth in the
case of IH is higher above approximately the average energy of the
original $\bar{\nu}_e$ spectrum\footnote{Note that the total
  luminosity is the same in all flavors.}.  Likewise, model A provides
a higher flux than model B at high energies due to its higher average
energies.

\begin{figure}[t]
\includegraphics[width=0.47\linewidth]{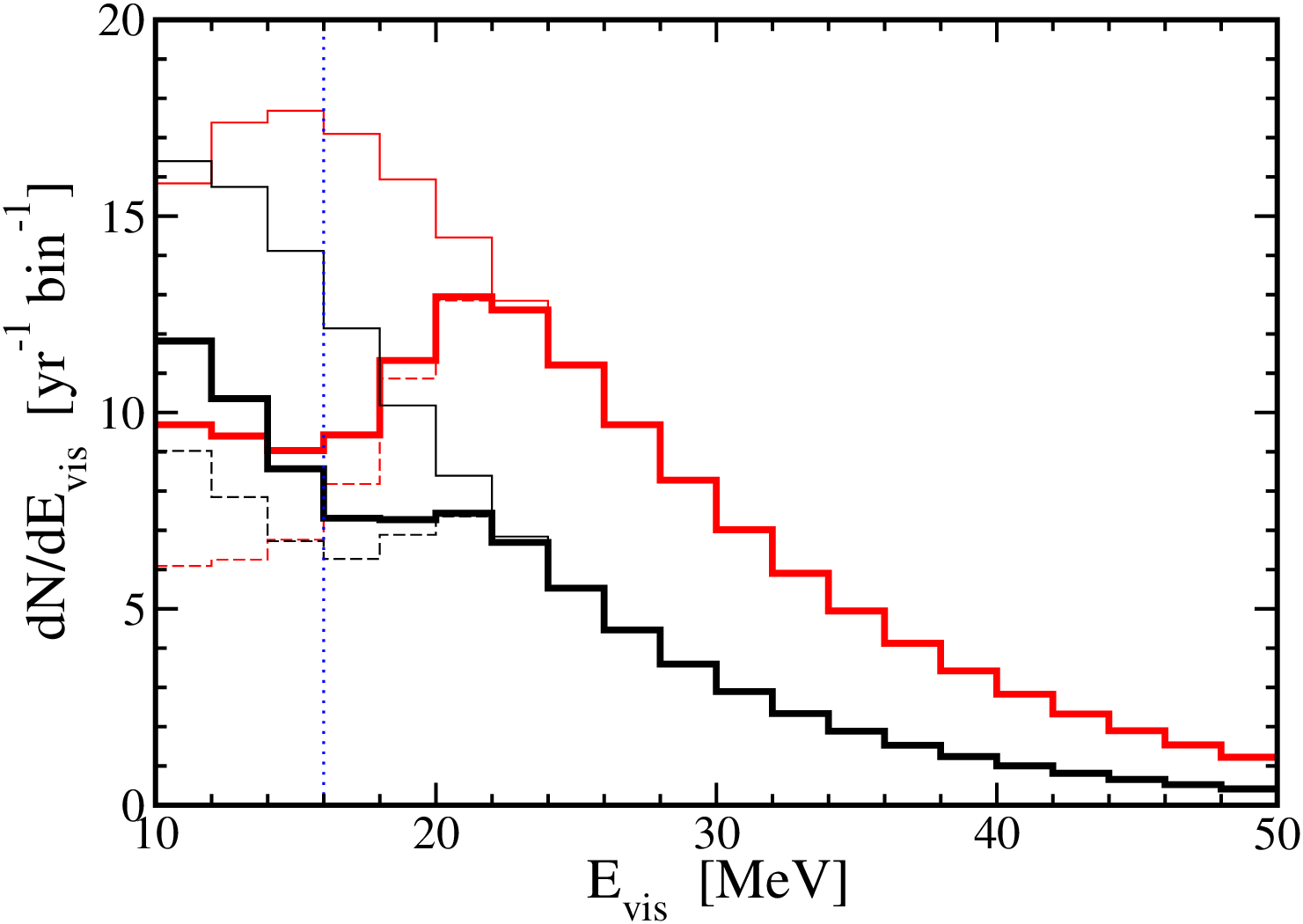}
\hspace{5mm}
\includegraphics[width=0.47\linewidth]{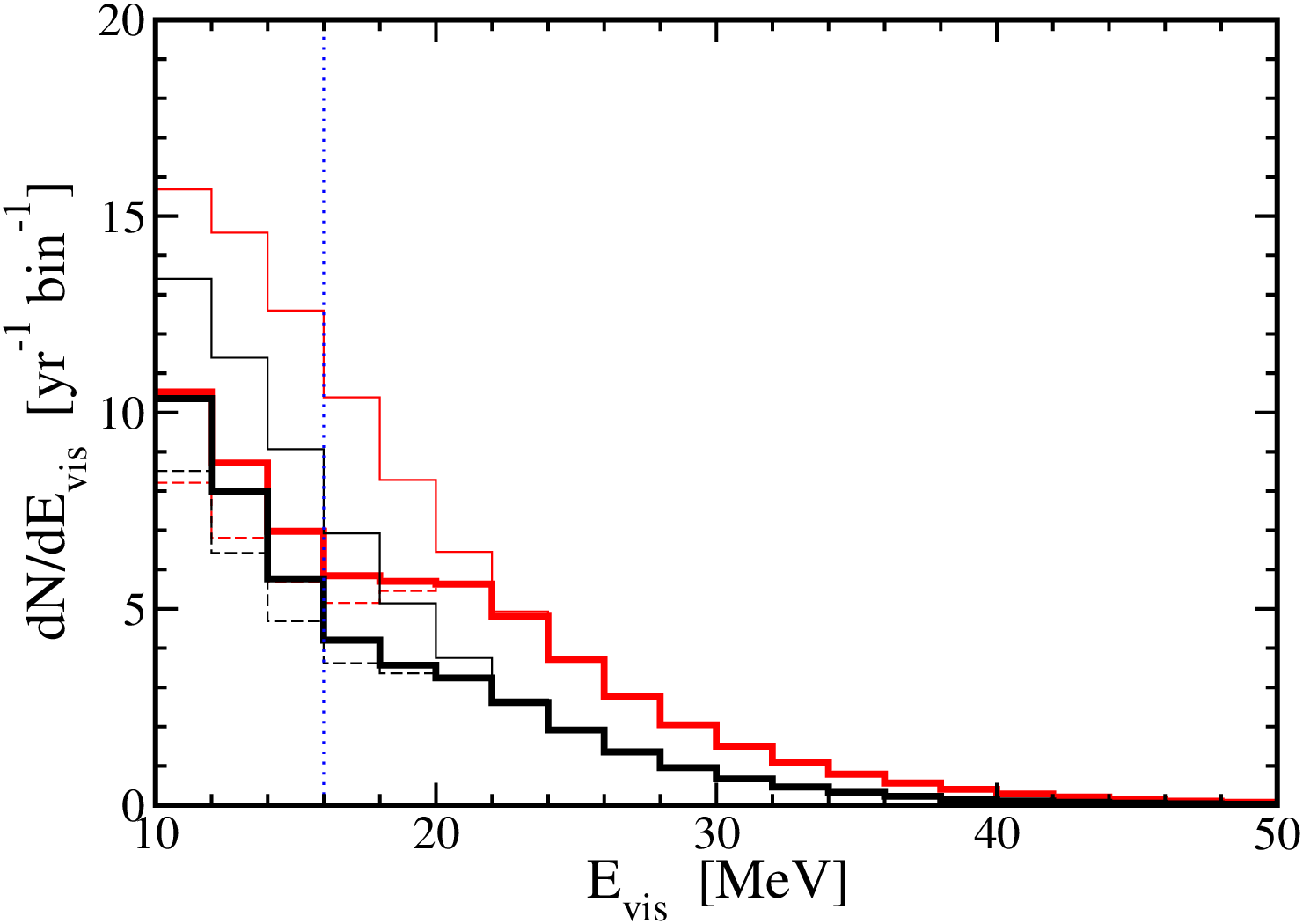} \\
\vspace{9mm}
\includegraphics[width=0.47\linewidth]{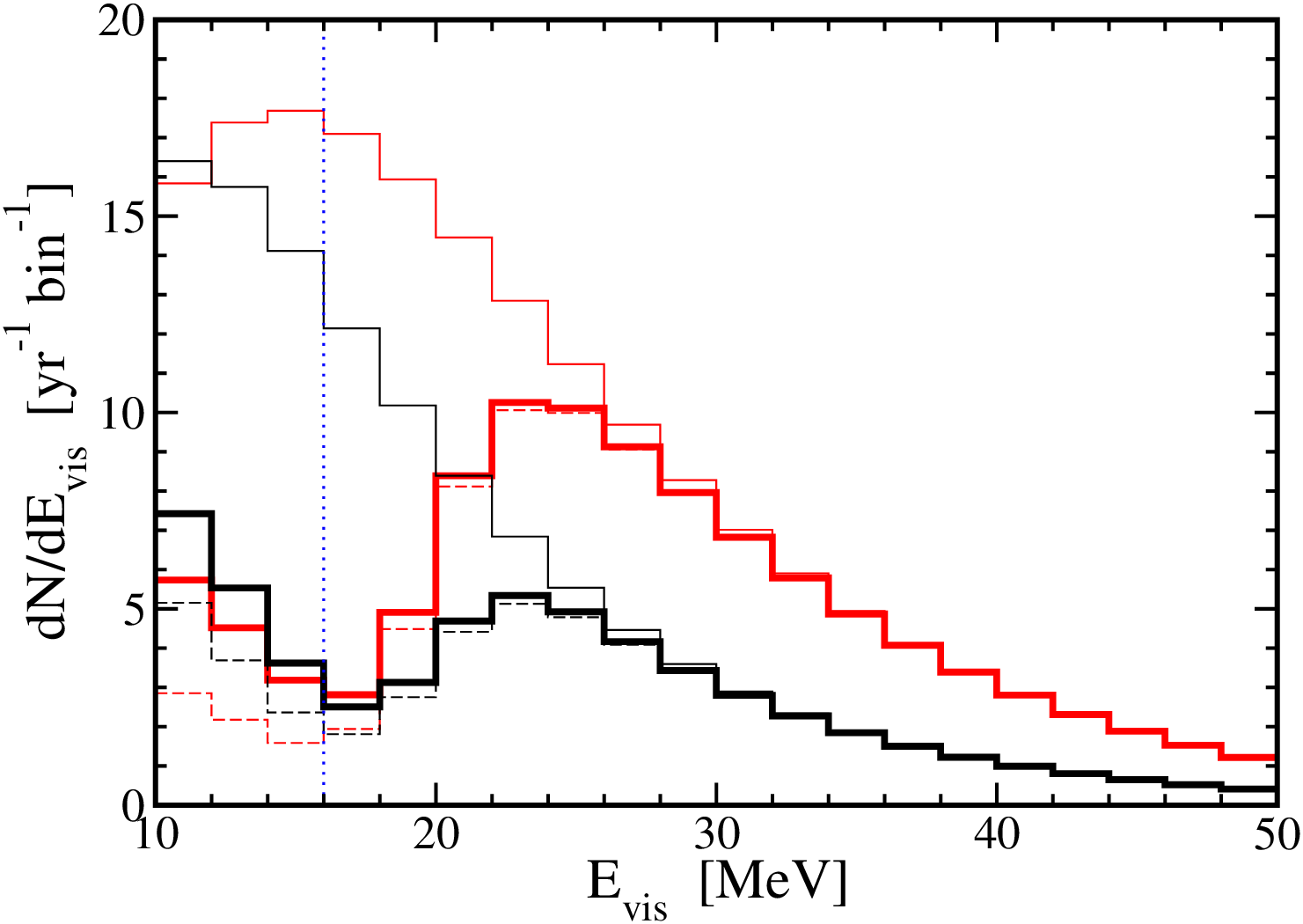}
\hspace{5mm}
\includegraphics[width=0.47\linewidth]{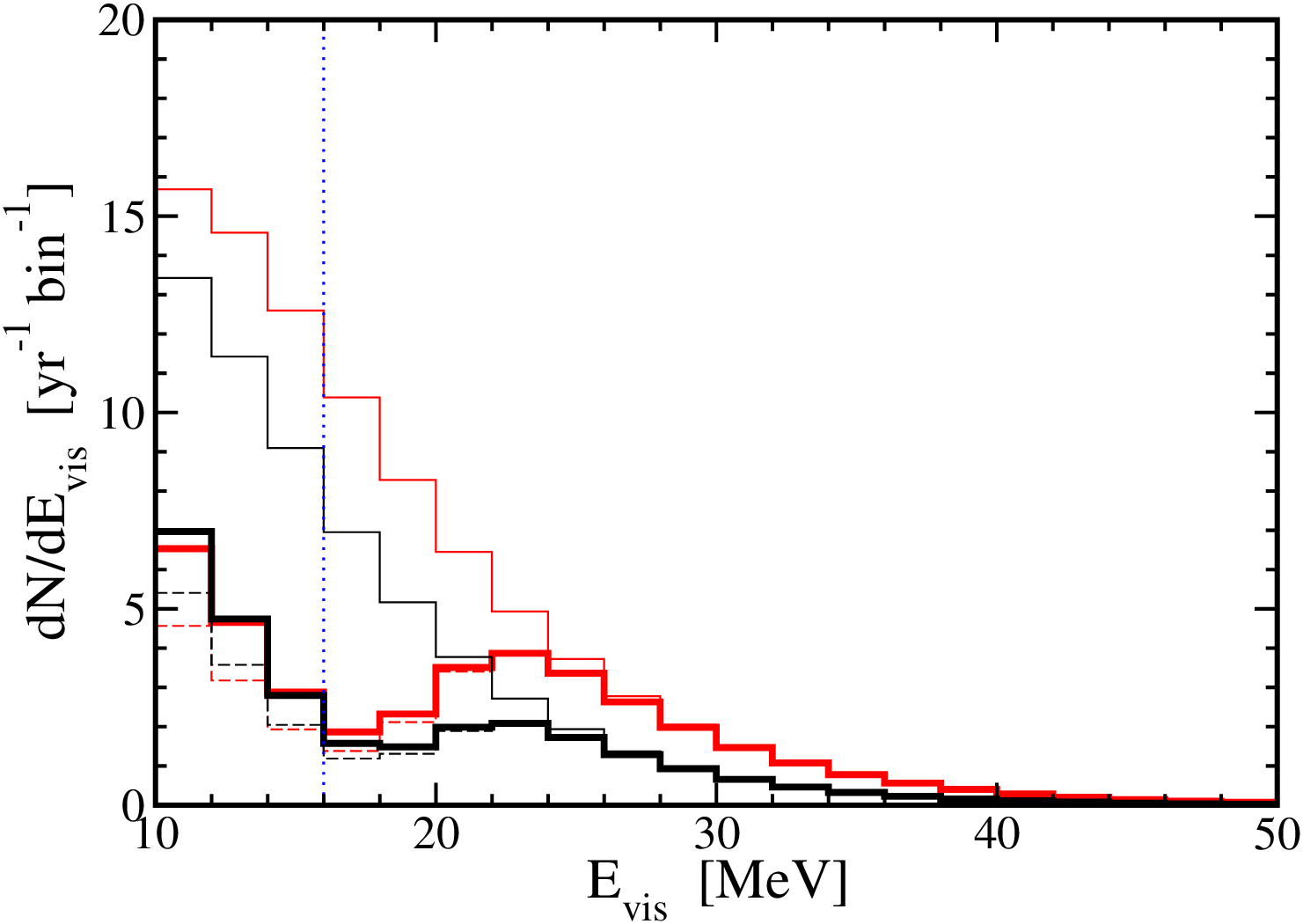}
\caption{{\sl Event spectra at HK for model A (left panels) and model B
    (right panels), assuming $g_\tau = 0.1$ (upper panels) and $g_\tau
    = 0.5$ (lower panels).  The common parameters and explanation of the
    curves are the same as those in Fig.~\ref{fig:spectra}.  We assume
    a fiducial volume of 562.5~kton ($25 \times$SK), a constant
    detection efficiency of 90\% and an energy resolution of 10\% over
    the whole visible (positron/electron) energy range.  The vertical
    dotted line is the current SK threshold.  $E_{\rm vis}$ represents
    the energy of the detected positron/electron.}}
\label{fig:events}
\end{figure}

Let us now consider the expected signal in a future water-\v{C}erenkov
detector like HK~\cite{Abe:2011ts}.  In Fig.~\ref{fig:events} we
depict the expected number of events per year in 2--MeV bins for the
cases shown in Fig.~\ref{fig:spectra}.  We assume a fiducial volume of
562.5~kton (25 times the volume used in the DSNB searches in SK), a
constant detection efficiency of 90\% and an energy resolution of 10\%
over the whole visible (positron/electron) energy range, which are
typical SK parameters~\cite{Bays:2011si, Baysthesis, Hosaka:2005um,
  Cravens:2008aa, Abe:2010hy}.  Although the main signal comes from
inverse beta decay events ($\bar{\nu}_e + p \to e^++n$), we also add
the contribution from the interactions of $\nu_e$ and $\bar{\nu}_e$
off Oxygen nuclei, as done in Refs.~\cite{PalomaresRuiz:2007eu,
  PalomaresRuiz:2007ry, Bernal:2012qh}.  In Fig.~\ref{fig:events},  the
dotted vertical line indicates the current SK energy threshold, which
is mainly set by the large number of spallation products.  However,
adding Gd would make the neutron tagging of the inverse beta decay
neutrons efficient~\cite{Beacom:2003nk}, which could be used to
significantly reduce backgrounds and move the detection threshold to
lower energies. Thus, we extend the event spectra down to 10~MeV.

We see that for $g_{\tau}=0.5$ (lower panels), in all the cases, even
for the SK threshold of 16~MeV, the drop in the event spectrum below
the resonance energy is very significant.  On the other hand, for
$g_{\tau}=0.1$, in the case of IH and model A, the dip is also very
likely to be detectable (even once backgrounds are properly added and
a full analysis performed).  This would signal the presence of new
physics producing a suppression with respect to the expected DSNB
flux.  For the other cases and $g_{\tau}=0.1$, the suppression of the
expected flux is also significant.  However it might be non trivial to
disentangle this signal from a spectrum with lower average energies.
For this less favorable case ($g_{\tau}=0.1$) and the assumed
parameters, we expect a $\sim$25\% ($\sim$15\%) effect in the whole
energy range considered in Fig.~\ref{fig:events} for a threshold
energy of 10~MeV (16~MeV), whereas for $g_{\tau}=0.5$, we expect a
suppression of $\sim$50\% ($\sim$40\%).  On the other hand, the
suppression when only considering the bins affected by the resonant
interaction is $\sim$35\% ($\sim$30\%) for $g_{\tau}=0.1$, and
$\sim$65\% ($\sim$60\%) for $g_{\tau}=0.5$.  In the less optimistic
scenario, i.e., $g_{\tau}=0.1$, model B and NH, HK would detect
$\sim$11 ($\sim$35) events/year in the energy interval $16~{\rm
  MeV}<E_{\rm vis}<22~{\rm MeV}$ ($10~{\rm MeV}<E_{\rm vis}<22~{\rm
  MeV}$) as compared to the $\sim$16 ($\sim$50) events/year when no
absorption occurs.  In the most optimistic scenario, i.e.,
$g_{\tau}=0.5$, model A and IH, HK would detect $\sim$16 ($\sim$30)
events/year as compared to the $\sim$47 ($\sim$98) events/year when
there is no absorption.  Hence, if a resonance of this type occurs in
the relevant energy range, after a few years, it could be possible to
determine the presence of the redshift-integrated resonance dip with
a reasonable confidence level.  Let us note that a detailed
statistical analysis including all the relevant backgrounds for
different experimental setups (with or without Gd) is beyond the scope
of this work.

Let us finally mention that in a liquid scintillator detector like the
proposed LENA~\cite{Wurm:2011zn}, tagging inverse beta decay events
is possible by the delayed coincidence between the prompt positron and
the gamma-ray from neutron capture.  Therefore, the only irreducible
backgrounds are reactor and atmospheric $\bar{\nu}_e$, and hence, LENA
would have a low energy threshold ($\sim$10~MeV) and twice the number
of protons of SK (about ten times fewer than HK).  In addition, LENA
would also be sensitive to all flavors via neutral-current reactions
as neutrino-proton or neutrino-electron elastic scattering, although
the statistics for these channels is at least an order of magnitude
lower, making the corresponding rate of the DSNB very low for the
study of the spectral features discussed in this work.

In the examples considered so far, we have taken $g_e=g_\mu=0$ because
for light $\phi$ and $N$, there are strong bounds on these couplings
from kaon decay data, below which the distortion of the DSNB spectrum is
negligible.  However, as discussed before for $m_\phi+m_N>m_K$, the
bounds from kaon decays on $g_e$ and $g_\mu$ can be relaxed.  For
such masses, to have a resonance energy below $\sim 30$ MeV, a fine
tuned cancelation between $m_\phi^2$ and $m_N^2$ is needed (see
Eq.~\rf{eq:Eres}).  In addition, the optical depth at resonance goes as
$g^2/(\Eres \, \mdm^2)$ and hence the magnitude of the dip is very
much suppressed.  We have studied the distortion of the spectrum for
$\mphi = \mdm = 240$~MeV, $\mn = \mr = 260$~MeV (for which
$\Eres=20.8$~MeV) and $g_e=g_\mu=0.4$ and found that the effect of the
interaction is negligible.

\section{Conclusions}
\label{sec:conclusions}

We have studied the distortion of the energy spectrum of DSNB within
scenarios with DM, with mass in the MeV range, coupled to ordinary
neutrinos $\nu$ as $g N_R^\dagger \nu_L \phi$ with the lighter new
particle (a scalar $\phi$ or a heavy neutrino $N$) playing the role of
DM.  We have found that such a coupling could give rise to a resonance
scattering of neutrinos off the ambient DM background.  Although the
resonance would be in general very narrow, the cumulative effect of
resonance scattering of neutrinos at different redshifts could lead to a
significantly wide dip in the spectrum at the detectors.  In order to
have a sizeable effect, the coupling should be relatively large $g >
{\rm few} \times 10^{-2} (\Eres/20 \, {\rm MeV})^{1/2} (\mdm/{\rm
  MeV})$.  The overall results are the same for all eight cases with
$m_N<m_\phi$ or $m_\phi>m_N$; $\phi$ being real or complex and $N$
being (pseudo-)Dirac or Majorana.  We have however focused on the case
with real $\phi$ and (pseudo-)Dirac $N$, for this is the only case for
which the annihilation cross section of the DM pair remains below the
thermal limit (i.e., $\sim$1~pb) even for couplings as large as $g\sim
{\cal O}(1)$.  Nevertheless, relaxing the the constraints on the
couplings imposed by the thermal scenario, other lepton number
conserving cases with large couplings could also be possible.

Strong upper bounds on the $g_e$ and $g_\mu$ couplings are also
imposed from studies of meson decays.  The bounds apply for all the
eight cases mentioned above, provided the sum of the masses of $N$ and
$\phi$ to be below the kaon mass.  In order to avoid these bounds, we
have focused on the following case: coupling exclusively to
$\nu_\tau$, i.e., $g_e=g_\mu=0$ and $g_\tau \ne 0$).  We have also
considered the case $g_e \ne 0; \, g_\mu \ne 0$, when $m_N+m_\phi
>m_K$ with $E_R<30$ MeV (see Eq.~\rf{eq:Eres}), although the expected
dip is very small.  For the lepton number violating case (Majorana $N$
and real $\phi$), the coupling induces a contribution to neutrino mass
at one-loop level so the upper bound on active neutrino masses can be
translated into a strong upper bound on the coupling.  We have shown
that in the limiting case of pseudo-Dirac $N$ and real $\phi$, the
active neutrino mass matrix could be reconstructed at one-loop level
despite the relatively large coupling giving rise to the distortion of
the energy spectrum of the DSNB.

When $\Eres \sim 20$~MeV and $g>0.1$, the distortion of the spectrum is
quite sizeable. The ability of a detector to establish the presence of
the dip depends on the statistics, and hence, on the detector mass and
energy threshold.  We have studied two models for the SN neutrino
spectra (see Tab.~\ref{tab:tab1}).  As is well known, the statistics
also depends on the neutrino mass hierarchy.  For IH, the number of
$\bar{\nu}_e$ events would be higher.  We have found that for the
favorable case of model A and IH, the presence of the dip could be
established after a few years of data taking by the upcoming HK
detector, clearly indicating the role of new physics at play.  For the
less favorable case of NH, the deviation of the spectrum from the
prediction of a specific SN model could be established, but the
distortion could be hidden within the uncertainties on the SN model,
i.e., because of the statistical errors, the effect of the dip in the
spectrum of the observed events could be reproduced by shifting the
average neutrino energies to lower values.  Nevertheless, a detailed
statistical analysis including all the backgrounds and
parametrizations of the SN neutrino spectra is beyond the scope of
this work.  We have also discussed the prospects of the proposed LENA
experiment, although its smaller size would render spectral analyses
challenging.

If the presence of the dip is established, such a feature would
indicate the presence of new physics.  The question would be then
whether one could establish the particular scenario and coupling we
have assumed.  The shape of the dip could be indicative of a resonance
scattering {\it en route} and could not be reproduced by other new
physics scenarios such as pseudo-Dirac neutrino
oscillation~\cite{Beacom:2003eu, Esmaili:2012ac}.  However, other
scenarios giving rise to resonance scattering (like scattering on the
background relic neutrinos via an s-channel light $Z^\prime$
exchange~\cite{Goldberg:2005yw, Baker:2006gm}) could give rise to a
similar feature.  Even with low energy experiments such as those
studying the meson or lepton decays, distinguishing between the two
scenarios would be challenging because they predict a similar
deviation from the SM expectations.  In order to distinguish between
these two cases, one should look for the signatures of the ultraviolet
complete model embedding these scenarios.

\section*{Acknowledgments}

We thank T.~Weiler for a careful reading of the manuscript.  SPR is
supported by a Ramón y Cajal contract and by the Spanish MINECO under
grant FPA2011-23596.  YF and SPR are also partially supported by the
European Union FP7 ITN INVISIBLES (Marie Curie Actions,
PITN-GA-2011-289442).  SPR was also partially supported by the
Portuguese FCT through the projects CERN/FP/123580/2011,
PTDC/FIS-NUC/0548/2012 and CFTP-FCT Unit 777
(PEst-OE/FIS/UI0777/2013), which are partially funded through POCTI
(FEDER).  SPR gratefully acknowledges the hospitality and financial
support of the Institute for Research in Fundamental Sciences (IPM),
where parts of this work were done.

\section*{Appendix: DM annihilation cross section}

In this section, we study the annihilation of dark matter pair via
the coupling in Eq.~\rf{eq:coupling}. The mass term for $N_R$ can in
general be written as
\ba{eq:NrNl}
(N_R^T \ N_L^T)c \left( \begin{matrix} m_R & m_D \cr m_D &
  m_L \end{matrix} \right) \left( \begin{matrix}N_R \cr
N_L \end{matrix}\right)  ~.
\ea

We denote the mass eigenvalues by $\mn$.  In the case $m_R,m_L=0$ ($\ll
m_D$), $N_R$ is a (pseudo-)Dirac fermion.  Otherwise, $N_R$ is a
Majorana fermion and  the presence of $N_L$ is unnecessary.  Lepton
number is conserved if $\phi$ is complex and/or $N$ is of Dirac type.

In the pseudo-Dirac limit, mass eigenstates are almost degenerate
Majorana fermions.  We denote the two mass eigenstates by
\be{eq:N1N2}
N_1\simeq \frac{(1+\alpha)N_L+(1-\alpha)N_R}{\sqrt{2}} \
\ {\rm and} \ \   N_2\simeq
i\frac{(1-\alpha)N_L-(1+\alpha)N_R}{\sqrt{2}} ~,
\ee
with $\alpha=(m_L-m_R)/2m_D$.  The corresponding mass eigenvalues are
\be{eq:M1M2}
m_{N_1}=m_D+\frac{m_L+m_R}{2}  \ \  {\rm and}
\ \   m_{N_2}=m_D-\frac{m_L+m_R}{2} ~.
\ee
Notice that substituting $N_R$ with
$((1-\alpha)N_1+i(1+\alpha)N_2)/\sqrt{2}$ in Eq.~\rf{eq:coupling}, we
find that the couplings of $N_1$ and $N_2$ are respectively equal to
$g/\sqrt{2}$ and $ig/\sqrt{2}$.

Let us now discuss the annihilation of DM pairs for the eight possible
cases mentioned in the text.

\begin{itemize}

\item {\it Case} $m_N<m_\phi$: In this case $N$ is the DM candidate.
  Let us discuss the four possible subcases one by one:

\begin{itemize}

\item {\it Real $\phi$ and Dirac $N$:} In this LC case we have
  \be{11}
  \sigma(N \, N\to \nu \, \nu) = \sigma(\bar{N} \, \bar{N} \to
  \bar{\nu} \, \bar{\nu}) = \frac{g^4}{4\pi }\frac{ \mn^2}{(\mn^2 +
    \mphi^2)^2} ~.
  \ee
  Moreover, the $N \bar{N}\to \nu \bar{\nu}$ annihilation mode is an
  s-wave and is also given by
  \be{eq:NNbar}
  \langle\sigma(N \bar{N} \to \nu \bar{\nu})v \rangle =
  \frac{g^4 \mn^2}{4\pi (\mn^2 + \mphi^2)^2} ~.
\ee

\item {\it Real $\phi$ and Majorana $N$:}  In this case, lepton number
  is violated and a pair of $N$'s can annihilate into $\nu \nu$ with the
  annihilation cross section given by Eq.~\rf{11}.  On the other hand,
  LC annihilation into $\nu \bar{\nu}$ is p-wave suppressed and
  therefore subdominant (see Eq.~\rf{eq:lightest}).

\item {\it Complex $\phi$ and Dirac $N$:} The LV annihilation into
  $\nu\nu$ or $\bar{\nu}\bar{\nu}$ pairs is forbidden, but LC pair
  annihilation into $\nu \bar{\nu}$ is allowed with a s-wave cross
  section given by Eq.~\rf{eq:NNbar}.

\item {\it Complex $\phi$ and Majorana $N$:} The dominant annihilation
  mode is the p-wave suppressed LC annihilation into a $\nu \bar{\nu}$
  pair:
  \be{eq:lightest}
  \langle\sigma(N {N} \to \nu \bar{\nu}) v\rangle=\frac{4 \,
    g^4}{3\pi}\frac{\mn^4 + \mphi^4}{(\mn^2 + \mphi^2)^4} \; p_{\rm
    DM}^2 ~,
\ee
where $p_{\rm DM}$ is the momentum of the DM at freeze-out: $p_{\rm
  DM}^2\sim m_N^2/20$.

\end{itemize}

Let us now comment on the pseudo-Dirac scenario with the mass
eigenstates $N_1$ and $N_2 $ and small mass splitting $\Delta \mn =
m_L+m_R$.  In early universe, these particles scatter off $\phi$ and
$\nu$ with a rate given by $\Gamma_{\rm scat}\sim g^4T/4\pi$.  At
scattering, the final state is the chiral $N_R$, which is a coherent
combination of $N_1$ and $N_2$.  As long as $\Delta \mn$ is small
enough to have $\Delta \mn /\Gamma_{\rm scat}\ll 1$, coherence between
mass eigenstates $N_1$ and $N_2 $ is preserved, so DM pairs would
interact with each other as Dirac particles.  In the opposite case of
$\Delta \mn /\Gamma_{\rm scat}\gg 1$, the coherence is lost and the $N_1$
and $N_2$ pairs annihilate and coannihilate with each other as
Majorana particles.

If we want to restrict to the thermal DM scenario, the total
annihilation cross section should be ${\cal O}(1)$~pb.  Thus, in all
cases with $N$ as thermal DM, the values of the couplings should be
$g<{\cal O}(0.01)$.

\item  {\it Case $m_\phi<m_N$:}
In this case, which is extensively studied in Refs.~\cite{Boehm:2006mi,
Farzan:2009ji, Farzan:2011tz}, $\phi$ is the DM candidate. Let us
consider the four different possibilities mentioned in Sec. II. If
$N$ is of Majorana type, regardless of whether $\phi$ is real or
complex, the LV annihilation channel $\phi \phi \to \nu \nu $ can take
place.  For real $\phi$, we  have $\sigma (\phi \phi \to \nu \nu)=\sigma
(\phi \phi \to \bar{\nu}\bar{\nu})$ while for complex $\phi$, we
have $\sigma(\phi \phi \to \nu \nu)=\sigma (\bar{\phi} \bar{\phi}
\to \bar{\nu}\bar{\nu})$.  If we restrict the possibility to the
thermal DM scenario, the total annihilation cross section should be
${\cal O}(1)$~pb, so the sum of these annihilation cross sections cannot
exceed this value.  Taking $m_L=0$, $m_D\ll m_R$ (see
Eq.~\rf{eq:NrNl}) and $\mn\simeq m_R\sim 1-10$~MeV, from
$\sigma(\phi\phi\to \stackrel{(-)}{\nu}\stackrel{(-)}{\nu})\sim 1$~pb,
it was found in Refs.~\cite{Boehm:2006mi, Farzan:2009ji, Farzan:2011tz}
that
\be{eq:SLIMrange}
3\times 10^{-4}<g<10^{-3} ~.
\ee
These LV modes are forbidden for Dirac $N$.  For the pseudo-Dirac
scenario, $\sigma (\phi \phi \to \nu \nu)$ is suppressed by
$(m_R/m_D)^2$.  As a result, by  taking $m_R/m_D$ arbitrarily small,
the upper bound in Eq.~\rf{eq:SLIMrange} can be avoided.  However,
for both Dirac and Majorana $N$, if $\phi$ is complex, there would be
p-wave annihilation into a $\nu \bar\nu$ pair: $\sigma(\phi
\bar{\phi} \to \nu \bar{\nu})\sim g^4 p_{DM}^2/[4\pi (\mphi^2 +
  \mn^2)^2]$.  Taking $p_{\rm DM}^2\sim \mphi^2/20$ at freeze-out, from
$\langle \sigma(\phi \bar{\phi} \to \nu \bar{\nu}) v \rangle\sim
1$~pb, we find $g\stackrel{<}{\sim} 0.01$.  As discussed in
Refs.~\cite{Boehm:2006mi, Farzan:2009ji, Farzan:2011tz}, for real
$\phi$ the cross section of annihilation into a $\nu \bar\nu$ pair is
extremely suppressed, so this annihilation channel does not constrain
significantly the coupling.  Thus, for (pseudo-)Dirac $N$ and real
$\phi$, there should be another coupling or mechanism to fix the DM
abundance to the observed value.
\end{itemize}

In summary, if $\phi$ plays the role of DM, the bounds from imposing
thermal production of DM (i.e., $\langle \sigma_{tot} v \rangle \sim
{\cal O}(1)$~pb) constrain the coupling to be smaller than $\sim {\cal
  O}(0.01)$, unless $\phi$ is real and $N$ is a (pseudo-)Dirac
fermion.  Of course, if the assumption of thermal production is
relaxed, this bound does not apply anymore.


\end{document}